\documentclass[a4paper,12pt]{article}
\usepackage{a4}
\usepackage{amsmath,amssymb}
\usepackage{epsfig}
\usepackage{color,blackdvi}
\usepackage{graphicx}

\newtheorem{axiom}{Spin Axiom}

\newcommand{\set}{\stackrel{!}{=}}      

\newcommand{\sd}{\hspace{.5mm}\cdot\hspace{.3mm}}

\newcommand{\vect}[1]{\underline{#1}}  
\newcommand{\spa}{\tilde s}
\newcommand{\spb}{\hat s}
\newcommand{\spc}{\check s}
\newcommand{\lct}{\eta} 
\newcommand{\norm}[1]{\frac{1}{c^{#1}}}

\title{Spin Axioms in Relativistic Continuum Physics}

\author{Heiko J. Herrmann, G. R\"uckner, W. Muschik, H.-H. v. Borzeszkowski\\ \small{Institut f\"ur Theoretische Physik}\\\small{Technische Universit\"at Berlin}\\\small{Hardenbergstr. 36}\\\small{D-10623 Berlin}}
\begin{document}
\bibliographystyle{unsrt}
\maketitle

\begin{flushright}
  \begin{minipage}{5cm}
{\small \textit{There is nothing so annnoying \\ as a good example}
    \begin{flushright}
      Mark Twain
    \end{flushright}}
\end{minipage}
\end{flushright}

\begin{abstract}\setlength{\parindent}{0em}\noindent
The 24 components of the relativistic spin tensor consist of $3+3$ basic spin fields and $9+9$ constitutive fields. Empirically only 3 basic spin fields and 9 constitutive fields are known. This empirem can be expressed by two spin axioms, one of them identifying 3 spin fields, and the other one 9 constitutive fields to each other. This identification by the spin axioms is material-independent and does not mix basic spin fields with constitutive properties. The approaches to the Weyssenhoff fluid and the Dirac-electron fluid found in literature are discussed with regard to these spin axioms. The conjecture is formulated, that another reduction from 6 to 3 basic spin fields which does not obey the spin axioms introduces special material properties by not allowed mixing of constitutive and basic fields.

\end{abstract}

\section{Introduction}
We investigate the constitutive theory of spin fluids in a relativistic context. General Relativity Theory (GRT) using Riemann geometry for geometrization of gravitation is currently assumed to be the most appropriate geometrization.\\
Here we consider in a relativistic  framework the general spin balance. The systematic reduction from 6 to 3 basic spin fields is introduced by two spin axioms which prohibit the mixing between the basic spin fields and the constitutive fields. This is obvious, because no special constitutive assumptions should be introduced by this reduction. As an example the Weyssenhoff fluid and the Dirac-electron fluid with regard to the general spin balance and the spin axioms are discussed.

\section{The Spin Balances}
Balance equations are differential equations for the wanted basic fields. Beside these fields constitutive quantities describing the material appear in the balances. This structural distinction into basic fields and constitutive properties is independent of writing down the balance equations in relativistic or non-relativistic form.\\
First of all we consider the non-relativistic spin balance for characterizing, what are the basic spin fields and what the constitutive properties.

\subsection{Non-relativistic spin balance}
Starting out with the definition of the angular momentum as a skew-symmetric 3-tensor of second rank ($i,j,\dots = 1,2,3$)
\begin{eqnarray}
  M^{ij} & = & x^{[i}\rho v^{j]} + \rho s^{ij}\label{one}
\end{eqnarray}
the balance  of angular momentum is
\begin{eqnarray}
\frac{1}{2} \partial_t (\rho s^{ij}) + \partial_t x^{[i} \rho v^{j]} + \partial_k \left( \frac{1}{2} \rho s^{ij} v^k - m^{ijk} + x^{[i} \rho v^{j]} v^k - x^{[i} t^{j]k} \right) & = & l^{ij} + x^{[i} f^{j]} \nonumber \\ \label{two}
\end{eqnarray}
Here the basic fields are (\ref{one}) consisting of the orbital $x^{[i}\rho v^{j]}$ and the spin angular momentum densities $\rho s^{ij}$. The constitutive properties are described in (\ref{two}) by the stress tensor $t^{jk}$ and the couple stress $m^{ijk}$. An external angular momentum is given by $l^{ij}$.\\
Because the specific spin $s^{ij} \Leftrightarrow s_k$ can be identified with the axial spin vector $s_k$ in $R^3$ the balance of angular momentum can be written down as a vector equation. If from this balance equation the balance of momentum multiplied by $\times \vect{x}$ is subtracted, we obtain the balance of spin in vector formulation \cite{MUPAEH01}
\begin{eqnarray}
\partial_t (\rho s^i) + \nabla_k \left(v^k \rho s^i - w^{i k} \right) + \epsilon^{i j k} t_{jk} & = & \rho g^i \label{z}
\end{eqnarray}
Here the distinction into the basic fields $\rho s$ and the constitutive quantities $w$ and $t$ is clear: There are 3 basic spin fields $s$, 9 fields of the couple stress $w$, and 3 fields from the balance of momentum by the skew-symmetric part $\epsilon : t$ of the stress tensor. A given external angular momentum density is $\rho g$.\\

\subsection{Relativistic spin balance}
The non-relativistic momentum flux density $\rho v^j$ in (\ref{one}) is replaced in relativistic formulations by the energy-momentum tensor $T^{\alpha\beta}$. Therefore the tensor order one of the momentum flux density is replaced by the order two of the energy-momentum tensor in a relativistic formulation. Consequently in special relativity theory the definition of skew-symmetric angular momentum becomes ($\alpha, \beta, \dots = 1,\dots,4$)
\begin{eqnarray}
M^{\alpha \beta \mu} & := & x^{[\alpha}T^{\beta]\mu} + S^{\alpha \beta \mu}\label{three}
\end{eqnarray}
and the balance results in
\begin{eqnarray}\label{angmo}
\partial_\mu M^{\alpha \beta \mu} & = & L^{\alpha\beta} + x^{[\alpha}f^{\beta]}
\end{eqnarray}
Here $S^{\alpha \beta \mu}$ is the skew-symmetric spin tensor and $L^{\alpha\beta}$ the external angular momentum. Caused by the extension of the tensor order in relativistic theories with respect to non-relativistic ones, we now have $6\times4\ =\ 24$ fields of the spin tensor, including the basic spin fields and the constitutive quantities. Of how many basic and constitutive fields these 24 fields are composed, can be seen, if the spin tensor is decomposed into its spatial and time-like parts by a 3+1 split. By use of the 4-velocity $u^\alpha$ the 3+1 split of the spin tensor is
\begin{eqnarray}
- S^{\sd\sd\nu}_{\lambda\mu} \ = \ S^{\sd\sd\nu}_{\mu\lambda} & = & s^{\sd\sd\nu}_{\mu\lambda} + \norm{2} \spa_{\mu\lambda} u^\nu + \norm{2} u_{[\mu} \spb^{\sd\nu}_{\lambda]} + \norm{2} \spc^{\sd\nu}_{[\mu|} u_{|\lambda]} + \norm{4} \spa_{[\mu} u_{\lambda]} u^\nu + \norm{4} u_{[\mu} \spb_{\lambda]} u^\nu \label{split}\\ 
& = & s^{\sd\sd\nu}_{\mu\lambda} + \norm{2} \spa_{\mu\lambda} u^\nu + \norm{2} u_{[\mu} \Xi_{\lambda]}^{\sd\nu} + \norm{4} u_{[\mu} \Xi_{\lambda]} u^\nu \label{split2}
\end{eqnarray}
Here we have introduced the following abbreviations
\begin{eqnarray}
s_{\mu\nu}^{\sd\sd\lambda} & := & S_{\alpha\beta}^{\sd\sd\gamma} h^\alpha_\mu h^\beta_\nu h^\lambda_\gamma \qquad\textrm{couple stress}\label{cstr}\\
\spa_{\mu\nu} & := & S_{\alpha\beta}^{\sd\sd\gamma} h^\alpha_\mu h^\beta_\nu u_\gamma \qquad\textrm{spin density} \label{sden}\\
\spb^{\sd\lambda}_{\nu} & := & S_{\alpha\beta}^{\sd\sd\gamma} h^\beta_\nu h^\lambda_\gamma u^\alpha\\
\spc^{\sd\lambda}_{\mu} & := & S_{\alpha\beta}^{\sd\sd\gamma} h^\alpha_\mu u^\beta h^\lambda_\gamma\\
\spa_\mu & := & S_{\alpha\beta}^{\sd\sd\gamma} h^\alpha_\mu u^\beta u_\gamma\\ 
\spb_\nu & := & S_{\alpha\beta}^{\sd\sd\gamma} u^\alpha h^\beta_\nu u_\gamma\\
\Xi_{\nu}^{\sd\lambda} & := & \spb_{\nu}^{\sd\lambda} -
\spc_{\nu}^{\sd\lambda} \qquad\textrm{spin stress} \label{sstr}\\
\Xi_{\nu} & := & \spb_{\nu} - \spa_{\nu} \qquad\textrm{spin density vector}
\label{sdv}\end{eqnarray}\\
The spin balance equation determines the divergence of the spin tensor
\begin{eqnarray}
S^{\alpha\beta\mu}_{\sd\sd\sd ;\mu} & = & T^{[\alpha \beta]} + L^{\alpha\beta}\label{srsb}
\end{eqnarray}
Here $T^{[\alpha \beta]}$ is the skew-symmetric part of the energy momentum tensor which couples the spin balance to the balance of momentum. Beyond this source, there may be other external moments $L^{\alpha\beta}$ from (\ref{angmo}). We now decompose (\ref{srsb}) by its 3+1 split which results in two parts, the hh-part and the hu-part. Because of the skew-symmetry of the spin tensor the uu-part is zero. In more detail the hh-part of the 3+1 split results in 3 equations\\
\begin{eqnarray}
h^\mu_\alpha h^\nu_\beta S^{\sd\sd\lambda}_{\mu\nu\sd ;\lambda} & =& \underbrace{s^{\sd\sd\lambda}_{\mu\nu\sd ;\lambda} h^\mu_\alpha h^\nu_\beta}_{\Green{\spadesuit}} + \underbrace{\norm{2} \spa_{\mu\nu ; \lambda} u^\lambda h^\mu_\alpha h^\nu_\beta + \norm{2} \spa_{\mu\nu} h^\mu_\alpha h^\nu_\beta u^\lambda_{\sd ; \lambda}}_{\Green{\diamondsuit}} - \nonumber\\
& & - \underbrace{\norm{2} u_{[\mu} \Xi_{\nu]}^{\sd\lambda} \left( h^\mu_\alpha h^\nu_\beta \right)_{;\lambda}}_{\Green{\triangle}} - \underbrace{\norm{4} u_{[\mu} \Xi_{\nu]} u^\lambda \left( h^\mu_\alpha h^\nu_\beta \right)_{;\lambda}}_{\Green{\triangle}}\nonumber\\
& = & \underbrace{T_{[\mu\nu]} h^\mu_\alpha h^\nu_\beta}_{\Green{\sharp}} + \underbrace{L_{\mu\nu} h^\mu_\alpha h^\nu_\beta}_{\Green{\natural}} \label{hhp}
\end{eqnarray}
and the 3 equations of the hu-part becomes
\begin{eqnarray}
h^\nu_\gamma u^\mu S^{\sd\sd\lambda}_{\mu\nu\sd ;\lambda} & = & \underbrace{s^{\sd\sd\lambda}_{\mu\nu\sd ; \lambda} h^\nu_\gamma u^\mu - \norm{2} \spa_{\mu\nu} h^\nu_\gamma u^\mu_{\sd ; \lambda} u^\lambda}_{\Magenta{\triangle}} + \nonumber\\
& & + \underbrace{\frac{1}{2} \norm{2} \Xi_{\nu ; \lambda} u^\lambda h^\nu_\gamma + \frac{1}{2} \norm{2} \Xi_\nu h^\nu_\gamma u^\lambda_{\sd ;\lambda}}_{\Magenta{\diamondsuit}} + \underbrace{\frac{1}{2} \norm{2} u_{\mu} \Xi^{\sd\lambda}_{\nu\sd ;\lambda} u^\mu h^\nu_\gamma}_{\Magenta{\spadesuit}} \nonumber\\
& = & \underbrace{T_{[\mu\nu]} h^\nu_\gamma u^\mu}_{\Magenta{\sharp}} + \underbrace{L_{\mu\nu} h^\nu_\gamma u^\mu}_{\Magenta{\natural}} \label{hup}\\
~ \nonumber
\end{eqnarray}
Now we compare (\ref{hhp}) and (\ref{hup}) with the non-relativistic balance of spin (\ref{z})
\begin{eqnarray}
\underbrace{\partial_t (\frac{1}{2} \rho s^i) + \nabla_k (v^k \frac{1}{2} \rho s^i)}_{\Green{\diamondsuit}} - \underbrace{w^{i k}}_{_{\Green{\spadesuit}}}) & = & \underbrace{- \epsilon^{i k j} t_{k j}}_{\Green{\sharp}} + \underbrace{\rho g^i}_{\Green{\natural}}\label{zz}
\end{eqnarray}
The $\Green{\diamondsuit}$-terms in (\ref{zz}), (\ref{hhp}), and (\ref{hup}) are equivalent to the total time derivative, the $\Green{\spadesuit}$-terms belong to the couple stress, the $\Green{\sharp}$-terms to the skew-symmetric part of the energy-momentum tensor, and the $\Green{\natural}$-terms to the external angular momentum. The $\Green{\triangle}$-terms can be interpreted as coupling terms between the hh- and hu-part in which the other fields appear
respectively. Consequently we obtain balance equations for the spin density (hh-part) and for the spin density vector (hu-part) which are coupled to each other by the $\Green{\triangle}$-terms.\\
From (\ref{split2}), (\ref{hhp}), and from (\ref{hup}) we can see,  how the 24 components of the spin tensor are distributed on the basic fields and the constitutive quantities. The first term in (\ref{split2}), the couple stress (\ref{cstr}), is according to (\ref{zz}) a constitutive equation. Because the couple stress is space-like, it represents 9 fields. The second term in (\ref{split2}), the spin density (\ref{sden}), is according to (\ref{hhp}) a basic spin field having 3 components. According to (\ref{hhp}) and (\ref{hup}) there are other spin fields, the spin stress (\ref{sstr}) and the spin density vector (\ref{sdv}). Comparison of (\ref{hhp}) and (\ref{hup}) with (\ref{zz}) yields, that the spin density vector is an other basic spin field and that the spin stress is a constitutive quantity. The spin density vector field has 3 components, whereas the spin stress has 9 components. Consequently we have $3+3=6$ basic spin fields and $9+9=18$ constitutive quantities which all together result in the 24 components of the spin tensor.

\section{Spin Axioms}
As discussed in the previous section there are 6 basic spin fields in relativistic theories. But up to now only 3 spin fields are known in physics, a fact which is formulated in the following
\begin{quote}
{\bf Empirem}\\     
According to (\ref{zz}) in non-relativistic physics only 3 basic spin fields (the spin density) and 9 constitutive functions (the couple stress) are known and measurable.
\vspace{-.7cm}
\begin{flushright}{$\square$}\end{flushright}
\end{quote}
This empirem can be interpreted in two different ways: 3 of the 6 spin fields are caused by typically relativistic effects. They are very small and not be measured up to now. Interestingly these 6 spin fields are connected with 18 constitutive couple stresses so that also materials should have relativistic properties which cannot be detected in the non-relativistic limit. If this possibility of interpretation seems to be too artificial, the other possibility remains: 
\begin{quote}
\begin{axiom}~\\
Also in relativistic physics only 3 basic spin fields and 9 couple stresses exist.
\end{axiom}
\vspace{-.7cm}
\begin{flushright}{$\square$}\end{flushright}
\end{quote}
A consequence of axiom 1 is to reduce the 6 spin fields to 3 ones in relativistic theories. The question is how to reduce them. \\
One first possibility is to demand, that the spin tensor is totally skew-symmetric
\begin{equation}
S_{\alpha\beta\gamma}\ \set S_{[\alpha\beta\gamma ]}\label{tss}
\end{equation}
In this case we obtain only 4 spin fields which may correspond to the wanted 3 basic spin fields and 1 additional constitutive field. Therefore by (\ref{tss}) the constitutive equations are severely restricted by setting 8 of the 9 fields of the couple stress to zero.\\
The 3+1 split of the totally skew-symmetric spin tensor (\ref{tss}) is according to (\ref{split2})
\begin{eqnarray}
S_{[\mu\nu\lambda]} & = & s_{[\mu\nu\lambda]} + \norm{2} \spa_{[\mu\nu} u_{\lambda]} + \norm{2} u_{[\mu} \Xi_{\nu\lambda]} \label{tss-split}
\end{eqnarray}
with
\begin{eqnarray}
\Xi_\lambda & = & S_{[\mu\nu\gamma]} u^\gamma \left( h^\mu_\lambda u^\nu - h^\nu_\lambda u^\mu \right) \ = \ 0
\end{eqnarray}
Introducing
\begin{eqnarray}
  \Sigma_{\mu\nu} & = & \spa_{\mu\nu} + \Xi_{\mu\nu} \label{tss-xi}
\end{eqnarray}
(\ref{tss-split}) becomes
\begin{eqnarray}
S_{[\mu\nu\lambda]} & = & s_{[\mu\nu\lambda]} + \norm{2} \Sigma_{[\mu\nu} u_{\lambda]}\label{tss-split1}
\end{eqnarray}
As discussed above $s_{[\mu\nu\lambda]}$ is the one constitutive field and $\Sigma_{\mu\nu}$ the three spin fields.\\

A second possibility of reducing fields is to cancel 3 spin fields and 9 constitutitive quantities arbitrarily. But this possibility seems not to be very systematic, because the basic spin fields and the constitutive fields are properly separated from each other: There is the spin density (\ref{sden}) and the spin density vector (\ref{sdv}) as basic spin fields, and the couple stress (\ref{cstr}) and the spin stress (\ref{sstr}) correspond to each other as constitutive fields. Because the hh-part (\ref{hhp}) and the hu-part (\ref{hup}) of the spin balance are coupled to each other, a third possibility
\begin{eqnarray} 
\Xi_{\nu}\ =\ 0, \quad & \wedge & \quad \Xi_{\nu}^{\sd\lambda}\ =\ 0 \label{xi}\\
&\textrm{or}&\nonumber \\
\spa_{\mu\nu}\ =\ 0,\quad & \wedge & \quad s_{\mu\nu}^{\sd\sd\lambda}\ =\ 0 \label{es}
\end{eqnarray}
results in differential equations which are not of balance type any more. Inserting (\ref{xi}) into (\ref{hhp}) and (\ref{hup}) we obtain from (\ref{hhp}) a balance equation of the spin density, whereas (\ref{hup}) has to be interpreted as a constraint for the constitutive quantity couple stress: The couple stress cannot be chosen arbitrarily, because its divergence is restricted by the constraints (\ref{hup}). The same happens to the spin stress and the spin density vector, if (\ref{es}) is inserted into (\ref{hhp}) and (\ref{hup}). Consequently by choice of (\ref{xi}) or (\ref{es}) hidden material properties are introduced.\\
An essential point is, that the reduction of the fields should not restrict the free choice of the also reduced constitutive equations. Consequently the possibility of identifying spin density with spin density vector and couple stress with spin stress remains.
\begin{quote}
\begin{axiom}~\\
Spin density and spin density vector field are semi-dual to each other, as couple stress and spin stress are, that means we identify
\begin{eqnarray}
\spa_{\alpha\beta}  \set  \frac{1}{2} \norm{2} \lct_{\alpha\beta\delta\gamma} u^\delta \Xi^\gamma \label{sax1a} & \Leftrightarrow & \norm{2} u_{[\mu}\Xi_{\nu]} = \frac{1}{2} \lct_{\mu\nu}^{\sd\sd\alpha\beta} \spa_{\alpha\beta} \label{sax1b}
\end{eqnarray}
\begin{eqnarray}
s_{\alpha\beta\gamma} \set \frac{1}{2} \norm{2} \lct_{\alpha\beta\delta\lambda} u^\delta \Xi^{\lambda}_{\sd\gamma} \label{sax2a} & \Leftrightarrow & \norm{2} u_{[\mu}\Xi_{\nu]}^{\sd\lambda} = \frac{1}{2} \lct_{\mu\nu}^{\sd\sd\alpha\beta} s_{\alpha\beta}^{\sd\sd\lambda} \label{sax2b}
\end{eqnarray}
\begin{flushright}{$\square$}\end{flushright}
\end{axiom}
\end{quote}
Here $\lct_{\alpha\beta\delta\lambda}$ is the Levi-Civita symbol. From (\ref{sax1a}) we see that $\spa_{\alpha\beta}$ and $\Xi^\gamma$ are semi-dual to each other if $\spa_{\alpha\beta}$ and $u^\delta \Xi^\gamma$ are dual to each other. According to (\ref{sax2a})  the same is valid for $s_{\alpha\beta\gamma}$ and $\Xi^{\lambda}_{\sd\gamma}$.\\

Inserting (\ref{sax1b})$_2$ and (\ref{sax2b})$_2$ into (\ref{split2}) we obtain for the spin tensor
\begin{eqnarray}
S^{\sd\sd\lambda}_{\mu\nu}  & = &  \left( \delta^\alpha_{[\mu} \delta^\beta_{\nu]} + \frac{1}{2} \lct^{\sd\sd\alpha\beta}_{\mu\nu} \right) \norm{2} \spa_{\alpha\beta} u^\lambda + \left( \delta^\alpha_{[\mu} \delta^\beta_{\nu]} + \frac{1}{2} \lct^{\sd\sd\alpha\beta}_{\mu\nu} \right) s^{\sd\sd\lambda}_{\alpha\beta}\label{st1}
\end{eqnarray}
and inserting (\ref{sax1a})$_1$ and (\ref{sax2a})$_1$ into (\ref{split2}) results in
\begin{eqnarray}
S^{\sd\sd\lambda}_{\mu\nu}  & = & \left( \delta^\gamma_{[\mu} \delta^\delta_{\nu]} + \frac{1}{2} \lct^{\sd\sd\gamma\delta}_{\mu\nu} \right) \norm{4} \Xi_\gamma u_\delta u^\lambda + \left( \delta^\gamma_{[\mu} \delta^\delta_{\nu]} + \frac{1}{2} \lct^{\sd\sd\gamma\delta}_{\mu\nu} \right) \norm{2} \Xi_\gamma^{\sd\lambda} u_\delta \label{st2}
\end{eqnarray}
By adopting the spin axiom 2 (\ref{st1}) and (\ref{st2}) are different but equivalent representations of the spin tensor, (\ref{st1}) in spin density and couple stress, (\ref{st2}) in spin density vector and spin stress. From (\ref{st1}) and (\ref{st2}) we obtain
\begin{eqnarray}
S^{\sd\sd\lambda}_{\mu\nu} & = & \left( \delta^\alpha_{[\mu} \delta^\beta_{\nu]} + \frac{1}{2} \lct^{\sd\sd\alpha\beta}_{\mu\nu} \right) \left( \norm{2} \spa_{\alpha\beta} u^\lambda + s^{\sd\sd\lambda}_{\alpha\beta\sd} \right) \label{st1a} \\
& = & \left( \delta^\gamma_{[\mu} \delta^\delta_{\nu]} + \frac{1}{2} \lct^{\sd\sd\gamma\delta}_{\mu\nu} \right) \left( \norm{4} u_\gamma \Xi_\delta  u^\lambda + \norm{2} u_\gamma \Xi_\delta^{\sd\lambda}\right) \label{st2a}
\end{eqnarray}
The common bracket in the representations of the spin tensor has the remarkable property
\begin{eqnarray}
\frac{1}{2} \lct^{\sd\sd\mu\nu}_{\kappa\rho} \left( \delta^\alpha_{[\mu} \delta^\beta_{\nu]} + \frac{1}{2} \lct^{\sd\sd\alpha\beta}_{\mu\nu} \right) & = & \left( \delta^\alpha_{[\kappa} \delta^\beta_{\rho]} + \frac{1}{2} \lct^{\sd\sd\alpha\beta}_{\kappa\rho} \right)
\end{eqnarray}
Consequently we have proven the
\begin{quote}
\textbf{Corollary}\\
By the spin axioms the spin tensor is self dual with respect to the first two indices
\begin{eqnarray}
S^{\sd\sd\lambda}_{\mu\nu} & = & \frac{1}{2} \lct^{\sd\sd\alpha\beta}_{\mu\nu} S^{\sd\sd\lambda}_{\alpha\beta}
\end{eqnarray}
\end{quote}
From the two representations (\ref{st1a}) and (\ref{st2a}) for the spin tensor we obtain two equivalent versions of the spin balance
\begin{eqnarray}
S^{\sd\sd\lambda}_{\mu\nu\sd ;\lambda} & = & \left( \delta^\alpha_{[\mu} \delta^\beta_{\nu]} + \frac{1}{2} \lct^{\sd\sd\alpha\beta}_{\mu\nu} \right) \left( \norm{2} \spa_{\alpha\beta} u^\lambda + s^{\sd\sd\lambda}_{\alpha\beta\sd} \right)_{;\lambda} \label{sbsa1}\\
& = & \left( \delta^\gamma_{[\mu} \delta^\delta_{\nu]} + \frac{1}{2} \lct^{\sd\sd\gamma\delta}_{\mu\nu} \right) \left( \norm{4} u_\gamma \Xi_\delta  u^\lambda + \norm{2} u_\gamma \Xi_\delta^{\sd\lambda}\right)_{;\lambda} \label{sbsa2}
\end{eqnarray}

The 3+1 decomposition of (\ref{sbsa1}) gives\\
\underline{\textbf{$h^\mu_\kappa h^\nu_\sigma$:}}
\begin{eqnarray}
h^\mu_\kappa h^\nu_\sigma S^{\sd\sd\lambda}_{\mu\nu\sd ;\lambda} & = & h^\mu_\kappa h^\nu_\sigma \left( \delta^\alpha_{[\mu} \delta^\beta_{\nu]} + \frac{1}{2} \lct^{\sd\sd\alpha\beta}_{\mu\nu} \right) \left( \norm{2} \spa_{\alpha\beta} u^\lambda + s^{\sd\sd\lambda}_{\alpha\beta\sd} \right)_{;\lambda}  \nonumber\\
& = & h^\mu_\kappa h^\nu_\sigma T_{[\mu\nu]} + h^\mu_\kappa h^\nu_\sigma L_{[\mu\nu]}\nonumber\\
& = & t_{[\kappa\sigma]} + h^\mu_\kappa h^\nu_\sigma L_{[\mu\nu]}\label{sbsa1hh}
\end{eqnarray}
\underline{\textbf{$u^\mu h^\nu_\sigma$:}}
\begin{eqnarray}
u^\mu h^\nu_\sigma S^{\sd\sd\lambda}_{\mu\nu\sd ;\lambda} & = & u^\mu h^\nu_\sigma \left( \delta^\alpha_{[\mu} \delta^\beta_{\nu]} + \frac{1}{2} \lct^{\sd\sd\alpha\beta}_{\mu\nu} \right) \left( \norm{2} \spa_{\alpha\beta} u^\lambda + s^{\sd\sd\lambda}_{\alpha\beta\sd} \right)_{;\lambda}  \nonumber\\
& = & u^\mu h^\nu_\sigma T_{[\mu\nu]} + u^\mu h^\nu_\sigma L_{[\mu\nu]}\nonumber\\
& = & \left( q_\sigma - p_\sigma \right) + u^\mu h^\nu_\sigma L_{[\mu\nu]}\label{sbsa1hu}
\end{eqnarray}

And the 3+1 decomposition of (\ref{sbsa2}) results in\\
\underline{\textbf{$h^\mu_\kappa h^\nu_\sigma$:}}
\begin{eqnarray}
h^\mu_\kappa h^\nu_\sigma S^{\sd\sd\lambda}_{\mu\nu\sd ;\lambda} & = & h^\mu_\kappa h^\nu_\sigma \left( \delta^\gamma_{[\mu} \delta^\delta_{\nu]} + \frac{1}{2} \lct^{\sd\sd\gamma\delta}_{\mu\nu} \right) \left( \norm{4} u_\gamma \Xi_\delta  u^\lambda + \norm{2} u_\gamma \Xi_\delta^{\sd\lambda}\right)_{;\lambda}  \nonumber\\
& = & h^\mu_\kappa h^\nu_\sigma T_{[\mu\nu]} + h^\mu_\kappa h^\nu_\sigma L_{[\mu\nu]} \nonumber\\
& = & t_{[\kappa\sigma]} + h^\mu_\kappa h^\nu_\sigma L_{[\mu\nu]}\label{sbsa2hh}
\end{eqnarray}
\underline{\textbf{$u^\mu h^\nu_\sigma$:}}
\begin{eqnarray}
u^\mu h^\nu_\sigma S^{\sd\sd\lambda}_{\mu\nu\sd ;\lambda} & = & u^\mu h^\nu_\sigma \left( \delta^\gamma_{[\mu} \delta^\delta_{\nu]} + \frac{1}{2} \lct^{\sd\sd\gamma\delta}_{\mu\nu} \right) \left( \norm{4} u_\gamma \Xi_\delta u^\lambda + \norm{2} u_\gamma \Xi_\delta^{\sd\lambda} \right)_{;\lambda}  \nonumber\\
& = & u^\mu h^\nu_\sigma T_{[\mu\nu]} + u^\mu h^\nu_\sigma L_{[\mu\nu]}\nonumber\\
& = & \left( q_\sigma - p_\sigma \right) + u^\mu h^\nu_\sigma L_{[\mu\nu]}\label{sbsa2hu}
\end{eqnarray}

Equations (\ref{sbsa1hh}) and (\ref{sbsa2hh}) are identical, as well as equations (\ref{sbsa1hu}) and (\ref{sbsa2hu}) are. The reduction from 6 to 3 equations (spin balances) requires, that also equations (\ref{sbsa1hh}) and (\ref{sbsa1hu}) are connected with each other, as well as equations (\ref{sbsa2hh}) and (\ref{sbsa2hu}) are. This leads to the requirement that also $(q_\sigma - p_\sigma)$ and $t_{[\sigma\kappa]}$ are semi-dual to each other:
\begin{eqnarray}
t_{[\kappa\sigma]} = \frac{1}{2} \norm{2} \lct_{\kappa\sigma}^{\sd\sd\beta\alpha}u_\beta (q_\alpha - p_\alpha) \label{const-rel}
\end{eqnarray}
As in SRT and GRT the spin balance is not connected with the field equations, $t_{[\kappa\sigma]}$ describes the coupling between the energy-momentum balance and the spin balance. According to spin axiom 2 this coupling requires (\ref{const-rel}) independently of the material under consideration. Therefore (\ref{const-rel}) is not a constitutive equation. In ECT the spn balance is connected to the field equations, and therefore the character of (\ref{const-rel}) has to be investigated separately.\\

The particular meaning of the spin axioms (\ref{sax1a}) and (\ref{sax2a}) can be characterized by the following proposition which we will prove elsewhere
\begin{quote}
{\bf Conjecture}\\  
All reductions from $6+18$ to $3+9$ basic and constitutive spin fields which do not use spin axiom 2 are introducing a specially chosen material.
\vspace{-.7cm}\begin{flushright}$\square$\end{flushright}
\end{quote}
If this conjecture is true, the reduction of the spin fields has to obey the spin axiom 2, because a reduction has to be performed before constitutive properties are introduced into the theoretical considerations.\\
With respect to the spin axioms we now will shortly discuss two examples of spin fluids well-known from the literature.

\section{Examples}
There are two spin fluids which are typically different from each other: The Weyssenhoff fluid which is a classical one, and the Dirac-electron fluid which represents the classical description of a quantum-fluid. We now will discuss these fluids with respect to the spin axioms introduced above.

\subsection{Weyssenhoff fluid}
According to \cite{WEYRA47,OBKO87} the spin tensor of the Weyssenhoff fluid reads:
\begin{eqnarray}
S^{\sd\sd \mu}_{\alpha\beta} & \dot = & \norm{2} \spa_{\alpha\beta}u^\mu \label{w1}
\end{eqnarray}
This choice of the spin tensor can be interpreted in two ways:
Comparing (\ref{w1}) with (\ref{st1}) we could state, that the couple
stress $s^{\sd\sd\lambda}_{\alpha\beta}$, the spin density vector
$\Xi_\nu$, and the spin stress $\Xi_\nu^{\cdot\lambda}$ are chosen 
to zero and that
therefore the choice (\ref{w1}) of the spin tensor does not satisfy
the spin axiom 2, because in (\ref{w1}) the $\eta$-term of the first
bracket in (\ref{st1}) is missing. Consequently we obtain from
(\ref{hhp}) the balance of the spin density
\begin{eqnarray}
\underbrace{\norm{2} \spa_{\mu\nu ; \lambda} u^\lambda h^\mu_\alpha h^\nu_\beta + \norm{2} \spa_{\mu\nu} h^\mu_\alpha h^\nu_\beta u^\lambda_{\sd ; \lambda}}_{\Green{\diamondsuit}}
= \underbrace{T_{[\mu\nu]} h^\mu_\alpha h^\nu_\beta}_{\Green{\sharp}} + \underbrace{L_{\mu\nu} h^\mu_\alpha h^\nu_\beta}_{\Green{\natural}} \label{w2}
\end{eqnarray}
and from (\ref{hup}) the constraint for the spin density
\begin{eqnarray}
\underbrace{- \norm{2} \spa_{\mu\nu} h^\nu_\gamma u^\mu_{\sd ; \lambda} u^\lambda}_{\Magenta{\triangle}} 
= \underbrace{T_{[\mu\nu]} h^\nu_\gamma u^\mu}_{\Magenta{\sharp}} + \underbrace{L_{\mu\nu} h^\nu_\gamma u^\mu}_{\Magenta{\natural}} \label{w3}
\end{eqnarray}
which only appears, if the spin axiom 2 is not taken into account.\\
If the spin axiom 2 is taken into account, (\ref{w1}) is compatible with (\ref{st1}), if
\begin{eqnarray}
\frac{1}{2 c^2} \lct^{\sd\sd\alpha\beta}_{\mu\nu} \spa_{\alpha\beta} u^\lambda + \left( \delta^\alpha_{[\mu} \delta^\beta_{\nu]} + \frac{1}{2} \lct^{\sd\sd\alpha\beta}_{\mu\nu} \right) s^{\sd\sd\lambda}_{\alpha\beta} & = & 0 
\label{w4}
\end{eqnarray}
or
\begin{eqnarray}
s^{\sd\sd\lambda}_{\mu\nu} & = & -\frac{1}{2} \lct^{\sd\sd\alpha\beta}_{\mu\nu} \left( \frac{1}{c^2} \spa_{\alpha\beta} u^\lambda + s^{\sd\sd\lambda}_{\alpha\beta} \right) \label{w5}
\end{eqnarray}
is valid. This is a constitutive equation of the couple stress which is not zero, if the spin axiom 2 is accepted. From (\ref{w5}) follows by the
\begin{quote}
{\bf Proposition}\\
\begin{eqnarray}
s^{\sd\sd\lambda}_{\mu\nu} & = & \norm{2} \left(\delta^\alpha_{[\mu} \delta^\beta_{\nu]} -  \lct^{\sd\sd\alpha\beta}_{\mu\nu} \right) \spa_{\alpha\beta}u^\lambda \label{w6}
\end{eqnarray}
\end{quote}
the couple stress as a function of the spin density.\\
Without accepting spin axiom 2 (\ref{w1}) describes an ideal spin fluid, but taking spin axiom 2 into account the spin fluid is not an ideal one, because the couple stress does not vanish according to (\ref{w6}).\\
 According to (\ref{st1}) the spin tensor of an ideal spin fluid is
\begin{eqnarray}
S^{\sd\sd\lambda}_{\mu\nu}  & = &  \left( \delta^\alpha_\mu \delta^\beta_\nu + \frac{1}{2} \lct^{\sd\sd\alpha\beta}_{\mu\nu} \right) \norm{2} \spa_{\alpha\beta} u^\lambda \label{w7}
\end{eqnarray}
if the spin axiom 2 is accepted.

\subsection{Dirac-electron fluid}
B\"auerle und Haneveld \cite{BAHA83} are starting out with a totally skew-symmetric spin tensor which is assumed to be dual to a not more specified spin vector
\begin{eqnarray}
S_{\mu\nu\lambda} & = & S_{[\mu\nu\lambda]} \ \dot = \ \lct_{\mu\nu\lambda\kappa} S^\kappa
\end{eqnarray}
This spin tensor consists of 3 independent components. One can rewrite it in the following form:
\begin{eqnarray}
 S^{\sd\sd\kappa}_{\mu\nu} g_{\kappa\lambda} & = & \norm{2} \tau_{[\mu\nu}u_{\lambda]} 
\end{eqnarray}
where $\tau$ is a spacelike skew-symmetric second rank tensor of 3 independent components. \\

As the spin tensor of B\"auerle und Haneveld is a special case of a totally skew-symmetric spin tensor, we will discuss this more general case.\\

The question arises if the totally skew-symmetric spin tensor (\ref{tss}), whose 3 + 1 split is (\ref{tss-split}), is compatible with the spin axioms.\\
First of all we obtain from (\ref{tss-xi}) and (\ref{sax1a}) that also $\spa_{\mu\nu}$ is zero
\begin{eqnarray}
\Xi_\lambda \ = \ 0 & \leftrightarrow & \spa_{\mu\nu} \ = \ 0
\end{eqnarray}
if spin axiom 2 is accepted. Therefore (\ref{tss-split}) results in
\begin{eqnarray}
S_{[\mu\nu\lambda]} & = & s_{[\mu\nu\lambda]} + \norm{2} \Xi_{[\mu\nu} u_{\lambda]} \label{tss-split2}
\end{eqnarray}
Here the four fields under consideration are $s_{[\mu\nu\lambda]}$ and $\Xi_{[\mu\nu]}$. By use of (\ref{sax2b})$_2$ we have
\begin{eqnarray}
\norm{2} u_{[\mu}\Xi_{\nu\lambda]} & = & \frac{1}{2} \lct_{[\mu\nu|}^{\sd\sd\alpha\beta} s_{\alpha\beta|\lambda]}
\end{eqnarray}
Consequently (\ref{tss-split2}) becomes
\begin{eqnarray}
S_{[\mu\nu\lambda]} & = & s_{[\mu\nu\lambda]} + \frac{1}{2} \lct_{[\mu\nu|}^{\sd\sd\alpha\beta} s_{\alpha\beta|\lambda]}
\end{eqnarray}
or in a shorter writing by introducing the bracket we obtain
\begin{eqnarray}
S_{[\mu\nu\lambda]} & = & \left( \delta^\alpha_{[\mu} \delta^\beta_{\nu|} + \frac{1}{2} \lct_{[\mu\nu|}^{\sd\sd\alpha\beta} \right) s_{\alpha\beta|\lambda]}
\end{eqnarray}
and only one spin field remains.\\
As clearly one spin field is not sufficient for describing spin effects, we can state that the choice of a totally skew-symmetric spin tensor is not compatible with the spin axioms. The reason for that is that the six spin fields $\Xi_\nu$ and $\spa_{\mu\nu}$ are zero. If we do not take the spin axioms into consideration, we obtain (\ref{tss-split1}) in which the spin fields $\spa_{\mu\nu}$ and the constitutive fields $s_{[\mu\nu\lambda]}$ and $\Xi_{\mu\lambda}$ are not properly separated, because $\Sigma_{\mu\lambda}$ appears. In \cite{BAHA83} the spin axioms are not accepted and the constitutive equations
\begin{eqnarray}
\Xi_{\mu\lambda} \ \dot = \ 0 & \textrm{and} & s{[\mu\nu\lambda]} \ \dot = 0 \label{bh2}
\end{eqnarray}
are presupposed, so that (\ref{tss-split}) results in
\begin{eqnarray}
S_{[\mu\nu\lambda]} & = & \norm{2} \spa_{[\mu\nu} u_{\lambda]} \label{bh1}
\end{eqnarray}
This is in accordance with (\ref{sax2a}), the second part of splin axiom 2, but does not satisfy part one (\ref{sax1a}).\\
As (\ref{w1}), (\ref{bh1}) contains exactly 3 spin ields in an ideal material (\ref{bh2}) (vacuum?), but in contrast to (\ref{w1}) (\ref{bh1}) cannot be adopted to the spiln axioms. Comparing the Weyssenhoff fluid (\ref{w1})  with the B\"auerle-Haneveld fluid (\ref{bh1}) we obtain without acceting the spin axioms

\begin{tabular}[c]{|l|c|c|c|c|c|}
\hline
& $S_{\mu\nu}^{\sd\sd\lambda}$ & $\spa_{\mu\nu}$ & $s_{\mu\nu\lambda}$ & $\Xi_{\mu\nu}$ & $\Xi_\nu$ \\
\hline
general & 24 or 4 & 3 & 9 or 1 & 9 or 3 & 3 or 0 \\
\hline
Weyssenhoff & 24 & 3 & $\dot =$ 0 & $\dot =$ 0 & $\dot =$ 0 \\
\hline
B\"auerle-Haneveld & 4 & 3 & $\dot =$ 0 & $\dot =$ 0 & $\equiv$ 0 \\
\hline
\end{tabular}

\vspace{.8cm}
{\bf Acknowledgement: }We thank F.W. Hehl for valuable hints concerning spin fluids and T. Chrobok for continuous discussions and co-working. Financial support by the DFG (German Research Council) and the Vishay Company, D-95085 Selb, Germany, is gratefully acknowledged.

\nocite{OBPI89}
\nocite{HHRUEMU00}
\nocite{RUEHHMU00}

\bibliography{spin}

\end{document}